\documentclass{PoS}
\usepackage{amsmath}
\usepackage{booktabs}
\usepackage{multirow}

\title{
Two-point Correlator Fits on HISQ Ensembles
}

\ShortTitle{
Two-point correlator fits on HISQ ensembles
}

\author{
A.~Bazavov,$^a$
C.~Bernard,$^b$
C.~Bouchard,$^c$
C.~DeTar,$^d$
D.~Du,$^e$
A.X.~El-Khadra,$^e$
J.~Foley,$^d$
E.D.~Freeland,$^f$
E.~Gamiz,$^g$
Steven~Gottlieb,$^h$
U.M.~Heller,$^i$
J.E.~Hetrick,$^j$
\speaker{J.~Kim},$^k$
A.S.~Kronfeld,$^l$
J.~Laiho,$^m$
L.~Levkova,$^d$
M.~Lightman,$^b$
P.B.~Mackenzie,$^l$
E.T.~Neil,$^l$
M.~Oktay,$^d$
J.N.~Simone,$^l$
R.L.~Sugar,$^n$
D.~Toussaint,$^k$
R.S.~Van~de~Water,$^{a,l}$
and
R.~Zhou$^h$
[Fermilab Lattice and MILC Collaboration]
\\
\llap{$^a$} Department of Physics, Brookhaven National Laboratory,\thanks{Operated by Brookhaven Science Associates, LLC, under Contract No.~DE-AC02-98CH10886 with the U.S. Department of Energy.} Upton, NY 11973, USA\\
\llap{$^b$} Department of Physics, Washington University, St. Louis, MO 63130, USA\\
\llap{$^c$} Department of Physics, The Ohio State University, Columbus, OH 43210, USA\\
\llap{$^d$} Physics Department, University of Utah, Salt Lake City, UT 84112, USA\\
\llap{$^e$} Physics Department, University of Illinois, Urbana,  IL 61801, USA\\
\llap{$^f$} Department of Physics, Benedictine University, Listle, IL 60532, USA\\
\llap{$^g$} CAFPE and Departamento de Fisica Te\,orica y del Cosmos, Universidad de Granada, Granda, Spain\\\llap{$^h$} Department of Physics, Indiana University, Bloomington, IN 47405, USA\\
\llap{$^i$} American Physical Society, One Research Road, Ridge, NY 11961, USA\\
\llap{$^j$} Physics Department, University of the Pacific, Stockton, CA 95211, USA\\
\llap{$^k$} Physics Department, University of Arizona, Tucson, AZ 85721, USA\\
\llap{$^l$} Fermi National Accelerator Laboratory,\thanks{Operated by Fermi Research Alliance, LLC, under Contract No.~DE-AC02-07CH11359 with the U.S. Department of Energy.} Batavia, IL 60510, USA\\ 
\llap{$^m$} SUPA, School of Physics and Astronomy, University of Glasgow, Glasgow G12 8QQ, UK\\
\llap{$^n$} Department of Physics, University of California, Santa Barbara, CA 93106, USA\\
E-mail: 
\email{jkim@physics.arizona.edu},
}

\abstract{We present our methods to fit the two point correlators for
light, strange, and charmed pseudoscalar meson physics with the highly
improved staggered quark (HISQ) action. We make use of the 
least-squares fit including the full covariance matrix of the correlators and
including Gaussian constraints on some parameters. We fit the
correlators on a variety of the HISQ ensembles. The lattice spacing
ranges from 0.15 fm down to 0.06 fm. The light sea quark mass ranges
from 0.2 times the strange quark mass down to the physical light quark
mass. The HISQ ensembles also include lattices with different volumes
and with unphysical values of the strange quark mass. We use the
results from this work to obtain our preliminary results of $f_D$,
$f_{D_s}$, $f_{D_s}/f_{D}$, and ratios of quark masses presented in
another talk \cite{Doug:2012}.}

\FullConference{The 30th International Symposium on Lattice Field Theory\\
     June 24-29,  2012\\
     Cairns, Australia}

\begin{document}

\section{Introduction}

Fitting correlators is one of the essential procedures in studying 
non-perturbative physics in Lattice QCD. The hadron spectrum and decay
constants can be calculated by fitting two-point correlation
functions, and hadronic form factor and matrix elements can be
calculated by fitting three-point functions. Here we present our
method for fitting two-point meson correlators with HISQ. 

The meson correlators are computed using the HISQ action\cite{Follana:2006rc}, 
which is an ${\cal O}(a^2)$-improved staggered quark action, with smaller
taste-symmetry breakings than in the asqtad action we
used previously.  
For the pseudoscalar meson correlators, we use both random-wall
sources and Coulomb gauge fixed wall sources,
with the point operator as the sink. We do
combined fits to the correlators of both sources. This helps us to
isolate the ground states that we are interested in. More details about
the correlators are given in Sec.~\ref{sec:cor}.

The HISQ lattices include the vacuum polarization of four dynamical
quarks: up, down, strange, and charm quarks
\cite{Bazavov:2010ru,Bazavov:2010pi}. There are, in total, nineteen HISQ
ensembles with different lattice spacings, quark
masses, and spatial volumes. The parameters of the HISQ lattices are
tabulated in Table \ref{tab:lat}. Note that we have physical sea-quark
mass ensembles, where all sea-quark masses are tuned to take the
physical values. These ensembles eliminate the need for a chiral 
extrapolation, and they can be used, together with higher unphysical-mass ensembles, 
to correct for slight mistunings of the sea quark masses.


Since elements of the correlators at different time slices are highly 
correlated, we use the full covariance matrix, which includes both 
diagonal and off-diagonal elements, to define the objective function 
$\chi^2$. We use Gaussian constraints, or priors in the language of Bayesian fits,
on some parameters of heavier states. This is 
useful especially when multiple states are needed to get reliable fits.

We divide the correlators into two groups: light-light and heavy-light. 
In this terminology, light refers to the up, down, and
strange quarks, while heavy refers to the charm quark. We find that we can
get good fits with a single state for light-light correlators 
as long as large enough distances are chosen for the fitting range. However, we 
find multiple state fits are efficient for heavy-light correlators to get 
good fits with small statistical errors. Our fitting methods are 
discussed in more detail in Sec.~\ref{sec:lsf}.

We block the correlators in the Monte Carlo trajectory to account 
for the autocorrelations. To find an optimal size of the blocks, 
we look at how the covariance matrix scales as the size of the blocks 
is increased. In the course of this analysis, we find that the 
autocorrelations depend on (Euclidean) time, and that 
they can be understood as a consequence of the way the
source time slices are chosen. A short discussion about this is given in
Sec.~\ref{sec:autocor}


The systematic error due to the excited-state contamination is tested by 
varying the fitting ranges and priors. Changes from these variations are small
compared to our statistical errors.  This is briefly discussed
in Sec.~\ref{sec:syserr} 

\section{Meson Correlators}
\label{sec:cor}

We use the point and wall operators to create the pseudoscalar mesons,
\begin{align}
  O_P(\vec{x}, t) &=
  \bar{\chi}_A (\vec{x}, t)
  \epsilon(\vec{x}, t) \chi_B (\vec{x}, t) \,,
  \label{eq:pop} \\
  O_W(t) &=
  \sum_{\vec{x}, \vec{y}} 
  \bar{\chi}_A (\vec{x}, t)
  \epsilon(\vec{x}+\vec{y}, t) \chi_B (\vec{y}, t) \,,
\end{align}
where $\chi$ is the HISQ field, and $A$, $B$ are flavor
indices. $\epsilon(\vec{x}, t) = (-1)^{\sum_i x_i + t}$
corresponds to the projection to the Goldstone pion with staggered quarks
 \cite{Golterman:1985dz}. The Coulomb gauge fixing must be used for the wall 
 operator, so we call this Coulomb wall operator.

Using the point sink, we can define two correlators
depending on the source,
\begin{align}
  C_{R}(t) &=  
  \Big\langle 
  \frac{1}{V_S}\sum_{\vec{x}} O_P(\vec{x},t)
  \frac{1}{3 V_S} \sum_{\vec{y}} O_P^{\dag}(\vec{y}, 0) \Big\rangle \,, \\
  C_{W}(t) &=  
  \Big\langle 
  \frac{1}{V_S}\sum_{\vec{x}} O_P(\vec{x},t)
  O_W^{\dag}(0) \Big\rangle \,.
\end{align}
The summation over the sources for the point operator is
implemented using random noise, so we call $C_R$ random wall source correlators. 

These correlators can be expressed in terms of masses and amplitudes
of relevant excitations. We adopt the following parameterization.
\begin{align}
  C_\text{R}(t) &= 
  \sum_{j=0}^{J-1} A_j M_j^3 \Big(e^{-M_j t} + e^{-M_j (N_T - t)}\Big) + 
  \sum_{k=0}^{K-1} (-1)^t A'_k {M'}_k^3 \Big( e^{-M'_k t} + e^{-M'_k (N_T - t)}\Big)  \,, \\
  C_\text{W}(t) &= 
  \sum_{j=0}^{J-1} B_j M_j^3 \Big(e^{-M_j t} + e^{-M_j (N_T - t)}\Big) + 
  \sum_{k=0}^{K-1} (-1)^t B'_k {M'}_k^3 \Big(e^{-M'_k t} + e^{-M'_k (N_T - t)}\Big) \,,
\end{align}
where $J$ and $K$ are the numbers of ordinary and alternate states respectively. The backward propagations from the image source at $N_T$,
the temporal size of the lattice, are included because of periodic 
boundary condition on the mesons.

Following \cite{Aubin:2004fs,Bazavov:2009bb}, we fit the two correlators simultaneously with common masses. It helps us to isolate the ground
states of the point source since the Coulomb wall operator is less contaminated from excited states. Once the masses and amplitudes are fitted,
the decay constant is given by 
\cite{Kilcup:1986dg}
\begin{align}
  af_{AB} = (am_A + am_B) \sqrt{3 V_S A_0/2} \,,
\end{align}
where $V_S$ is the spatial volume. Note that bare values of masses and amplitude can be used without renormalization thanks to the partially conserved axial current (PCAC) relation of HISQ. Hence, the uncertainty from renormalization is absent.

\begin{table}
\begin{center}
\begin{tabular}{ccccccc}
  $\beta$               &   $\approx a(fm)$                  & $am_l$             & $am_s$   & $L^3 \cdot T$    & $N_\text{lat}$ \\
  \midrule                                                                        
  \multirow{3}{*}{580}  &   \multirow{3}{*}{0.15}     & 0.013              & 0.065    & $16^3 \cdot 48$  & 1020 \\
                        &                            & 0.0064             & 0.064    & $24^3 \cdot 48$  & 1000 \\
                        &                            & 0.00235            & 0.064    & $32^3 \cdot 48$  & 1000 \\
  \midrule                                                                    
  \multirow{11}{*}{600} &   \multirow{11}{*}{0.12}    & 0.0102             & 0.0509   & $24^3 \cdot 64$  & 1040 \\
                        &                            & 0.00507            & 0.0507   & $24^3 \cdot 64$  & 1020 \\
                        &                            & 0.00507            & 0.0507   & $32^3 \cdot 64$  & 1000 \\
                        &                            & 0.00507            & 0.0507   & $40^3 \cdot 64$  & 1030 \\
                        &                            & 0.00184            & 0.0507   & $48^3 \cdot 64$  & 840\\
                        &                            & 0.00507            & 0.0304   & $32^3 \cdot 64$  & 1020 \\
                        &                            & 0.00507            & 0.022815 & $32^3 \cdot 64$  & 1020 \\
                        &                            & 0.00507            & 0.012675 & $32^3 \cdot 64$  & 1020 \\
                        &                            & 0.00507            & 0.00507  & $32^3 \cdot 64$  & 1020 \\
                        &                            & 0.00507/0.012675   & 0.022815 & $32^3 \cdot 64$  & 1020 \\
                        &                            & 0.0088725          & 0.022815 & $32^3 \cdot 64$  & 1020 \\
  \midrule                                                
  \multirow{3}{*}{630}  &  \multirow{3}{*}{0.09}     & 0.0074             & 0.037    & $32^3 \cdot 96$  & 1011 \\
                        &                            & 0.00363            & 0.0363   & $48^3 \cdot 96$  & 1000 \\
                        &                            & 0.0012             & 0.0363   & $64^3 \cdot 96$  & 702  \\
  \midrule                                               
  \multirow{2}{*}{672}  &  \multirow{2}{*}{0.06}    & 0.0048             & 0.024    & $48^3 \cdot 144$ & 1000 \\
                        &                           & 0.0024             & 0.024    & $64^3 \cdot 144$ & 655 \\
  \bottomrule
\end{tabular}
\end{center}

\caption{Parameters of HISQ ensembles are tabulated. With a few intentional 
exceptions, the light quark masses ($m_l$) are approximately 0.2, 0.1, and 
0.037 times the physical strange quark mass. The exceptions appear in the 
ensembles with unphysical values of the strange quark mass and non-degenerate 
up and down quark masses.
The HISQ ensembles also include three ensembles to study
finite-volume effects, where all parameters are the same except the
lattice volume. 
\label{tab:lat}}

\end{table}

\begin{table}
  \begin{center}
  \begin{tabular}{ccc}
  \toprule
          & light-light & heavy-light \\
  \midrule
  No. of states     &  J+K = 1+0        & J+K = 2+1 \\
  \midrule
  \multirow{2}{*}{Priors}   &
  \multirow{2}{*}{No}          & $M_1-M_0 = 700 \pm 140 $ MeV  \\
    &                          & $M'_0-M_0 = 400 \pm 200 $ MeV \\
      \midrule
  \multirow{1}{*}{Fit range} &
  \begin{tabular}{lll}
    $a$ (fm) & $t_\text{min}$  & $t_\text{max}$ \\
    0.15     & 16 & 23  \\
    0.12     & 20 & 31   \\
    0.09     & 30 & 47   \\
    0.06     & 40 & 71   \\
  \end{tabular}
  &
  \begin{tabular}{lll}
    $a$ (fm) & $t_\text{min}$  & $t_\text{max}$ \\
    0.15     & 8 & 23  \\
    0.12     & 10 & 31 \\
    0.09     & 15 & 47 \\
    0.06     & 20 & 71 \\
  \end{tabular} \\
  \bottomrule
  \end{tabular}
  \end{center}
  \caption{Our fit methods are summarized. $J+K$ stands for the
  number of states included in the fits, where $J$ $(K)$ is that of 
  ordinary (alternate) states. For heavy-light, the priors are
  applied to the mass gaps. The minimum distances are about
  2.4 fm for light-light and 1.2 fm for heavy-light. \label{tab:fit_sum}}
\end{table}

\section{Least-squares Fits}
\label{sec:lsf}

We use least-squares minimization to find the best fits. 
The objective function to be minimized is the augmented $\chi^2$ function,
which includes the Gaussian priors as well as usual $\chi^2$,
\begin{align}
  \chi^2_\text{aug} (\theta) = \chi^2(\theta) 
  + \sum_{\alpha}
  \frac{(\theta_\alpha - \mu_{\alpha})^2}{\sigma_{\alpha}^2} \,,
  \label{eq:chi2}
\end{align}
where $\theta$ represents a set of parameters to be estimated, and
$\alpha$ runs over parameters constrained with central value
$\mu_\alpha$ and width $\sigma_\alpha$.

Table \ref{tab:fit_sum} summarizes our fit methods. We divide the
correlators into two groups depending on the valence quark masses:
light-light and heavy-light, where by ``heavy'' we mean charm and
``light'', up, down, and strange. For light-light, we use
1+0 state with large minimum distances ($\sim 2.4$ fm). For
heavy-light, however, we use 2+1 state with constrained masses and
the reduced minimum distances ($\sim 1.2$ fm). We choose minimum distances 
small enough to provide adequate statistical signal for included states 
but not so small as to be polluted from ignored excited states.

\begin{figure}
\begin{center}
\includegraphics[width=0.54\textwidth]{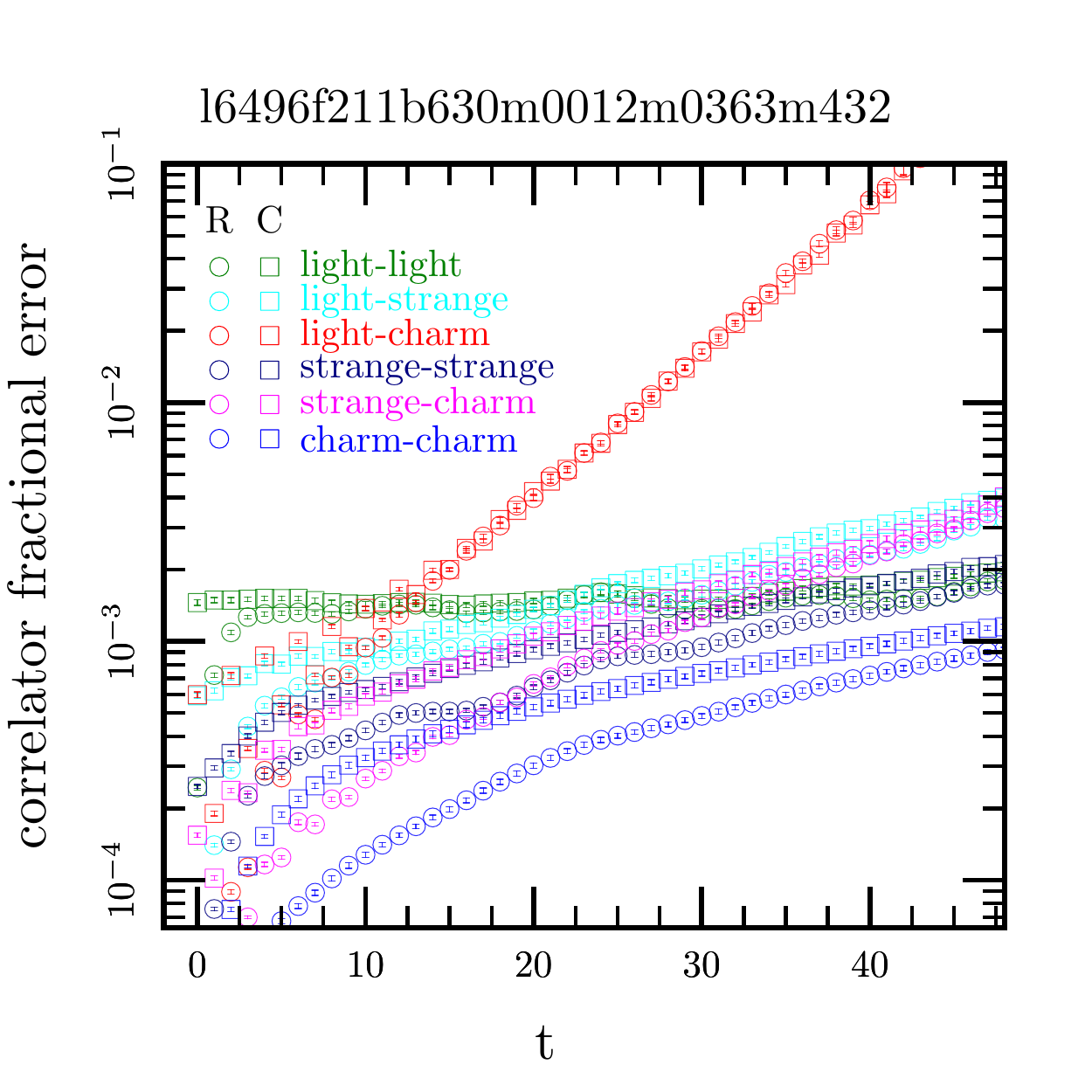}
\caption{A example of the fractional errors for various
valence quark mass combinations are shown. The circle and
square symbols stand for random and Coulomb wall sources
respectively. Here, ``light'' stands for up and down
quarks.\label{fig:fe}}
\end{center}
\end{figure}

The main reason for the differences in the two groups of fits is  the
fractional errors (noise-to-signal) of the correlators
\cite{Lepage:1989hd}. Figure \ref{fig:fe} shows an example of the
fractional errors for various valence quark combinations. One can
easily see that the fractional error grows much faster for heavy-light. 
This means that good signals are not expected at large
distances for heavy-light. Thus, it is better to reduce the minimum distance to get
better statistical precision, while including additional heavier
states in the fitting function.

The central values of the priors are taken to be 
$M_{D(2550)}-M_D \approx 700$ MeV for the ordinary excited
state and $M_{D^*}-M_{D} \approx 400$ MeV for the alternate state
\cite{pdg:2012}. The widths are chosen narrow enough to effectively
stabilize the fits, but not so narrow as to significantly influence our
determination of the ground states, the states of interest.
We find that these priors allow us to get good confidence levels of fits (or
$p$-values) over all heavy-light fits.


\begin{figure}
\includegraphics[width=0.5\textwidth]{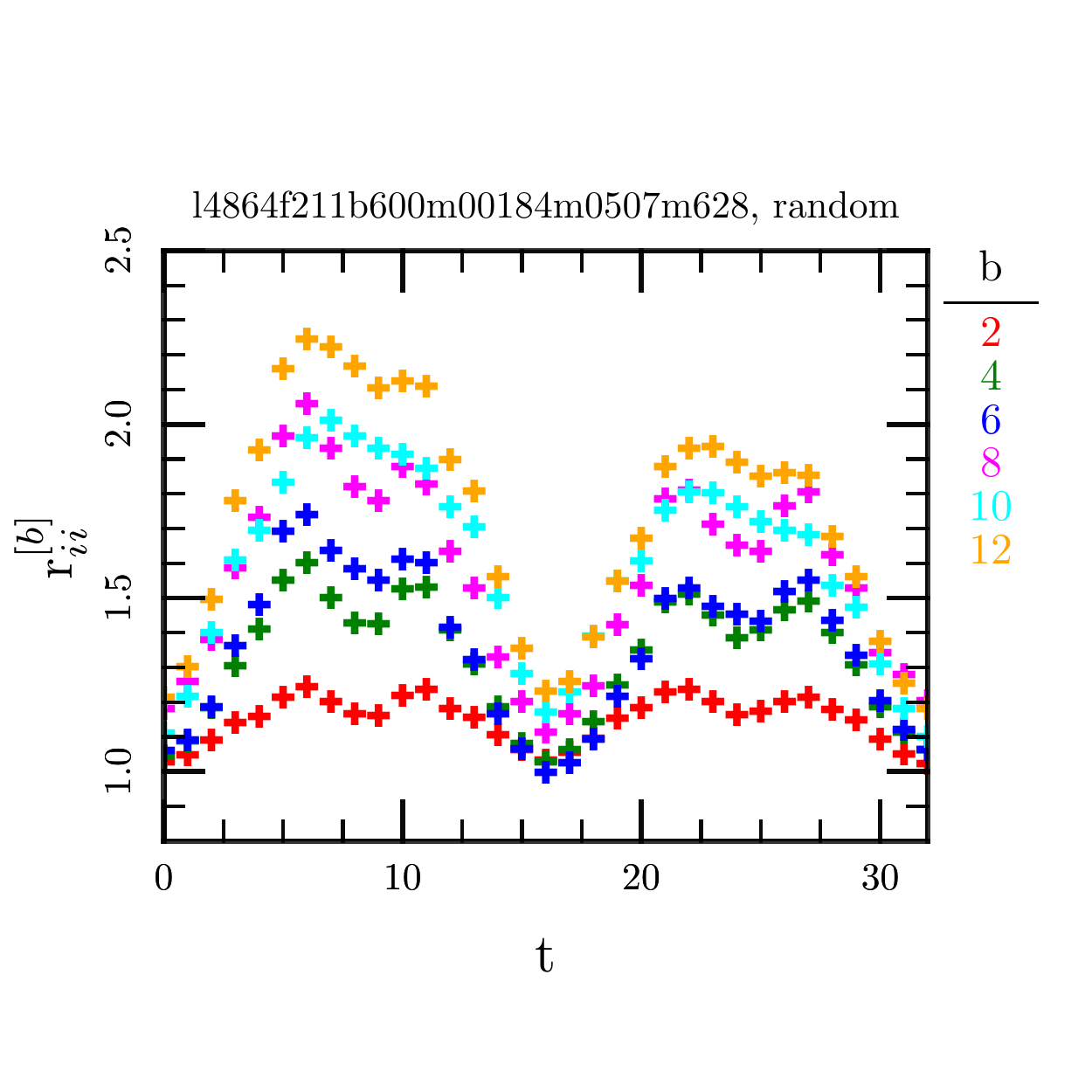}
\includegraphics[width=0.5\textwidth]{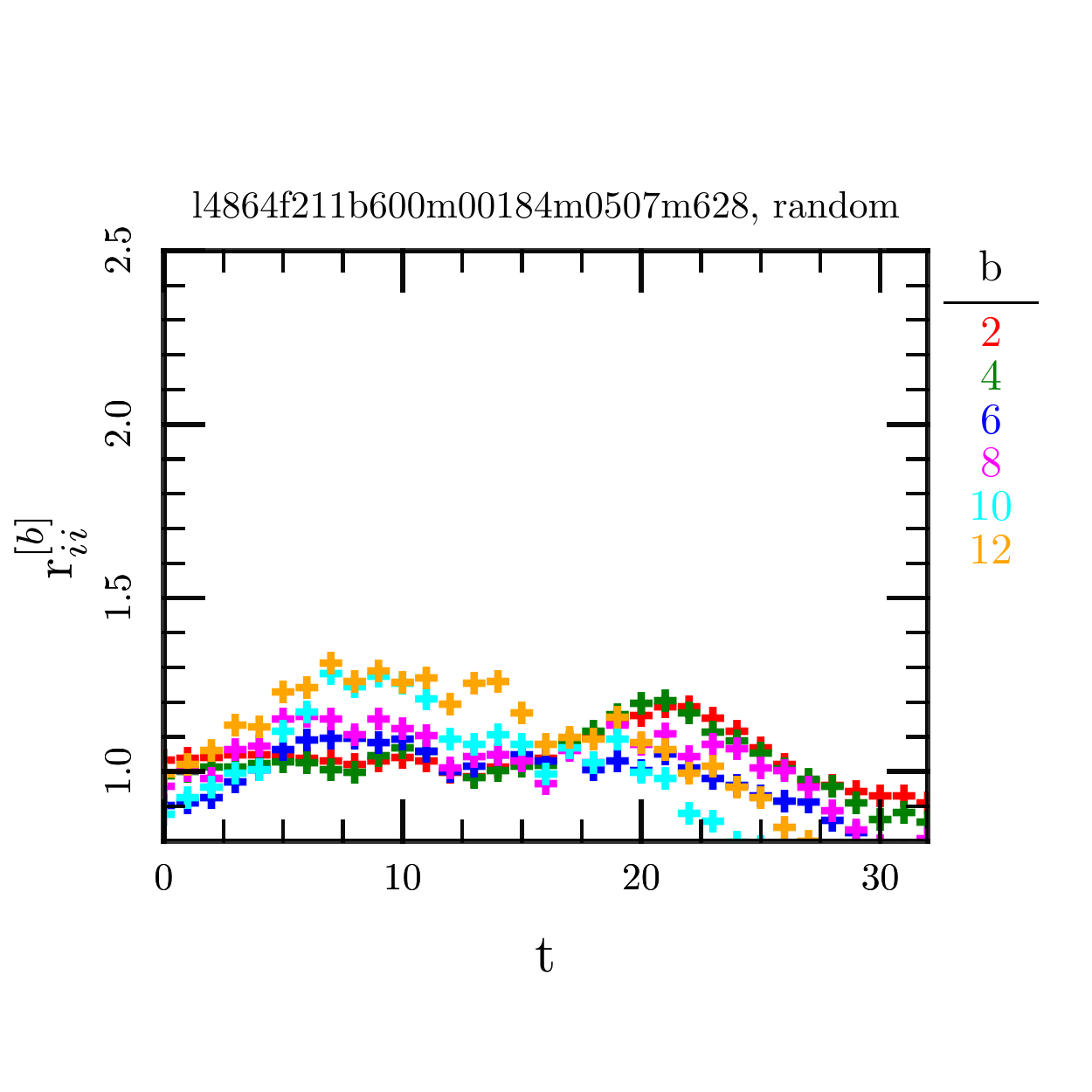}

\caption{The ratios of diagonal elements of the covariance matrix for
the random wall source correlator on the 0.12 fm physical quark mass
ensemble are plotted for various values of block size $b$. On the
left, the correlators are averaged over four measurements on each
lattice. On the right, however, only a single measurement is used to
measure the correlators. For details, please see the text. 
\label{fig:covratio}} \end{figure}

\section{Autocorrelation of Correlators}
\label{sec:autocor}

We use the blocking method to compute the correlators and their statistical
errors from the Monte Carlo samples. The blocking method divides measurements 
in a time series into contiguous blocks and uses their means as 
independent measurements in successive analysis. 
As a proxy for autocorrelation, we examine
ratios of the covariance matrix of the correlators as a function of the
block size,
\begin{align}
   r^{[b]}_{ij} \equiv C^{[b]}_{ij}/C^{[1]}_{ij} \,.
   \label{eq:covratio}
\end{align}
where $b$ denotes the size of blocks, and $C^{[b]}_{ij}$ represents the 
covariance matrix computed with the block size $b$.

Figure \ref{fig:covratio} shows examples of the diagonal elements of
the ratio matrix. On the left plot, we can see that the autocorrelation
has a peculiar (Euclidean) time dependence. This happens
because we measure the correlators from four source times and average them 
to get one estimator on each gauge configuration. Then, when we proceed
to the next gauge configuration, we move the locations of sources by
a constant amount $N_T/8$ plus a small displacement $\delta = \pm 1$,
which prevents moving back and forth to the same places. Note that the
positions of peaks coincide with the ``moving distance'', $N_T/8 +
\delta$, or its multiple. We can compare these with the results
of the single measurement (with the the moving distance $\sim N_T/4$), 
which are plotted on the right in Fig.~\ref{fig:covratio}. Hence,
we find these autocorrelations are induced
by the way we place the multiple sources thoughout the lattices.

To suppress the effects of the autocorrelations, we need to increase
the size of the blocks. However, as the block size increases we
have fewer blocks, and our error estimates get less accurate.
In this trade-off, we find that an optimal block size is four, which 
gives us enough statistics to estimate the covariance matrices of large dimension 
while reducing the underestimation of error due to the autocorrelations
for the ensembles where the number of configurations is about a thousand.  
We use the block size two or one depending on the number of configurations
when it is less than a thousand.

\section{Excited State Effects}
\label{sec:syserr}

To address the systematic uncertainty due to ignored excited states, we
vary the fitting range and priors and then test changes of the values
of interest. For measurements of the decay constants on the 0.09 fm 
physical sea quark mass ensemble, for example, no statistically significant deviation is
found beyond statistical precision with any of the following variations: with 
reduced prior widths ($\sim$ 50 MeV) for the mass gaps, with shifts of the central values
($\sim\pm$100 MeV), with additional constraints on the amplitudes, with
changes of the minimum distances ($\sim\pm$2), and so on. This
leads us to our conservative estimations of the systematic uncertainty 
from the excited states in Ref.~\cite{Doug:2012}.

\acknowledgments \vspace{-2.0mm}

This work was supported by the U.S. Department of Energy and National
Science Foundation.
Computation for this work was done at
the Texas Advanced Computing Center (TACC),
the National Center for Supercomputing Resources (NCSA),
the National Institute for Computational Sciences (NICS),
the National Center for Atmospheric Research (UCAR), 
the USQCD facilities at Fermilab,
and the National Energy Resources Supercomputing Center (NERSC),
under grants from the NSF and DOE.

\end{document}